\documentclass[useAMS,usenatbib, twocolumns]{mnras}
\usepackage{txfonts}
\usepackage{times}
\usepackage{latexsym, amssymb, amscd, psfrag}
\usepackage[dvipdfmx]{graphicx} 
\usepackage{bmpsize}
\usepackage{epsfig,color}
\usepackage{subfig}
\usepackage{morefloats,rotating,float}
\usepackage{gensymb}
\usepackage{multirow,array}
\usepackage{xcolor}
\usepackage{soul}
\usepackage{caption}
\usepackage{array,caption,threeparttable}
\usepackage[T1]{fontenc}
\usepackage{ae,aecompl}
\usepackage{multicol}
\usepackage{graphicx}    
\usepackage{amssymb}    
\usepackage{subfig}
\newcommand{\e}{{\sc eagle}}

\xdefinecolor{darkgreen}{rgb}{0.0,0.5,0.0}

\title[Dependence of RPS on the galactic orbit]{Study of dependence of ram pressure stripping on the orbital parameters of the galaxies} 
\author[Singh et al.]{Ankit Singh$^{1}$, Shreya Davessar ,$^{2}$
 Mamta Gulati,$^{2}$\thanks{E-mail: mamta.gulati@thapar.edu}
Jasjeet Singh Bagla,$^{3}$\thanks{E-mail: jasjeet@iisermohali.ac.in} \newauthor{Meenu Prajapati$^{2}$}
\\
$^{1}$ Korea Institute for Advanced Study (KIAS), 85 Hoegiro, Dongdaemun-gu, Seoul 02455, Republic of Korea\\
$^{2}$ Department of Mathematics, Thapar Institute of Engineering and Technology, Punjab 147004, India\\
$^{3}$ Department of Physical Sciences, Indian Institute of Science Education and Research (IISER) Mohali, \\
Knowledge City, Sector 81, Sahibzada Ajit Singh Nagar, Punjab 140306, India\\
}
\date{Accepted XXX. Received YYY; in original form ZZZ}
\pubyear{AAAA}

\begin{document}    
\captionsetup[subfigure]{labelformat=empty}
\renewcommand\thesubfigure{(\alph{subfigure})}
\label{firstpage}
\pagerange{\pageref{firstpage}--\pageref{lastpage}}
\maketitle

\begin{abstract}
Comprehensive observations of galaxy clusters suggest that gas deficiency in the galaxies could be due to ram pressure stripping due to the high-pressure intra-cluster medium acting on the galactic discs. The presence of gas in galaxies is essential for star formation. The net force due to ram pressure is dependent on the ambient medium and the orbit followed by the galaxy as it moves past the cluster medium. The current work deals with the effect of non-radial orbits of galaxies and the inclination of the disc plane of galaxies with the orbital plane on the mass of gas removed due to ram pressure. This gives a realistic approach to understanding the process of ram pressure stripping. The orbital parameters are extracted from \e~simulation data set along with the mass distribution of the galaxies. The analytical model proposed by Singh et. al. (2019) is modified appropriately to include the effect of the inclination angle. The non-radial orbits and infalling galaxies not being face-on decrease the amount of gas removed. Moreover, the inclination angle has a pronounced effect on the stripping of gas in low-mass galaxies as compared to high-mass galaxies with similar inclinations. The results show that the efficiency of the ram pressure stripping can be much lower in some cases, and hence gas in infalling galaxies can survive for much longer than expected from a simple analysis. 
\end{abstract}


\section{Introduction}

Galaxies form stars when gas collapses under the effect of gravity, eventually igniting nuclear fusion in the centres of protostars fragmenting within the cloud. The supply of cold gas determines the star formation rate in a galaxy. Observations have shown that the star formation rate in the universe attained a maximum between the redshift of $1-2$ and decreased to the present value with the decreasing redshift \citep[][and the references therein]{2014Madau}. Several mechanisms have been proposed to explain it, for example, AGN feedback, mergers, and environmental effects like harassment \citep{moore1998morphological}, strangulation \citep{larson1980evolution}, and ram pressure stripping \citet{gunn1972infall}. The importance of these mechanisms at different scales and under different scenarios has been a subject of research in recent years.

In the high-density environment, like groups or clusters, the galaxies close to the centre are observed to have very low HI gas \citep{cayatte1994very,cayatte1990vla,solanes2001hi,vollmer2006dynamical,van1990classification}, low star formation rate \citep{gomez2003galaxy,2021Mun}, early-type morphology \citep{goto2003morphology,pimbblet2003vigintennial} and high degree of assymetry in the galaxy disc \citep{Cramer_2023}. \citet{gunn1972infall} proposed a mechanism called Ram Pressure Stripping (RPS) to explain the dearth of gas in galaxies in clusters of galaxies. The galaxies move toward the cluster centre at high speeds and gas in these galaxies experiences pressure due to the high-pressure intra-cluster medium (ICM), leading to the dissociation of cold gas from galaxies (ram pressure). When the ram pressure on the gas exceeds the gravitational pull due to the remaining components of a galaxy, it gets stripped. The mass of the gas that is stripped depends upon the total stellar and gas mass of the galaxy, its trajectory about the cluster centre, the cluster mass, and the disk-wind angle.

Ram pressure stripping as a mechanism for galaxy quenching has been extensively studied in the literature \citep[][and the references within]{2021Boselli}. The study of the data from the Sloan Digital Sky Survey(SDSS) indicated that ram pressure is the dominant process for stripping as compared to slow processes like strangulation and galaxy harassment. More recently, \cite{2021Wang} using WALLABY survey that covers Hydra cluster showed that two-thirds of the galaxies residing inside the radius of $1.25$ $\times$ $R_{200}$ may be in the early stages of ram pressure stripping. Additionally, they reported an expected decrease of $0.1$ dex after the stripping-prone HI fraction is stripped. A recent study of jellyfish galaxy $ESO137-001$ \citep{li-yuan-luo-2023}, also suggested that the tail inherts the turbulent properties of ICM on interaction highlighting the role of RPS in its complex morphology and kinematics. Apart from the effect on star formation activity RPS has been linked to enhanced nuclear activity in the numerical simulations \citep{Akerman_2023,2020Ricarte}.

It is observed that gas from the galaxy is first removed from larger galactocentric distances, which favours RPS as the dominant mechanism, causing HI deficiencies in the galaxies \citep{schumacher2004ram, koopmann1998trouble}. The anemic galaxies (low HI surface densities) very close to the core of the cluster are observed to be even more gas-deficient \citep{giovanelli1983hi, van1976new, 2001Chengalur, 2009Lah}. In the Virgo cluster, observations have shown spiral galaxies to be HI deficient, and the HII regions in the outer disc are also absent. Cold gas is observed to be lying above the disc plane in some cases, suggesting that the molecular gas is being stripped from the disc and star formation occurs in this stripped gas  \citep{kenney1999ongoing, kenney2004vla, kenney2004spiral, 2017Schaefer}. The $H_{\alpha}$ trails in the intermediate-mass galaxies of the Coma cluster also indicated the dominant stripping process to be RPS \citep{gavazzi2018ubiquitous}. In a study of 19 spirals in the Coma cluster, the gas deficiency was observed in the outer parts of the galaxies \citep{bravo2000vla, clemens2000ram, boselli2006environmental}. This was attributed by the authors to interactions with the ICM, e.g., RPS. Comparison of stripping radii of Virgo cluster galaxies at different stages of RPS done by \citet{lee-2022} shows that the conventional RPS relation works reasonably well in a broad sense when RPS is the most dominant process and the galaxy is located where the surrounding environment can be well defined.  

Although most gas-deficient galaxies are closer to the core (within a distance of 0.6 Mpc from the centre), the environmental effect is not limited to the central cluster regions. The stripping of both hot and cold gas and, therefore, a decline in star-forming galaxies can be seen up to a distance as far as five times the virial radius from the cluster centre \citep{bahe2013does,gavazzi2006halpha}. At low redshift, in a group environment, the star formation rate is observed to be low for massive halos compared to the field galaxies \citep{treyer2018group}. 

In recent years many numerical and analytical techniques have been used to study the effects of various parameters on RPS {\citep{2021Vega, 2019Ayromlou, singh2019ram}. These studies have helped to improve our understanding of the physical processes needed to explain the observations. \citet{abadi1999ram} used Smoothed Particle Hydrodynamics(SPH)/N-Body simulation to show that galaxies with similar morphology passing through gas-rich clusters lose more gas due to RPS compared to gas-poor clusters. Mass loss due to RPS, calculated analytically by \citep{tecce2010ram}, leads to the conclusion that the stripping is due to ram pressure strongly affects the cold gas content in both small and massive clusters, and the rate of stripping depends on the virial mass of the
halo.

Previous studies have shown that the amount of stripped gas from the infalling galaxy reduces if the infalling galaxy is inclined \citep{abadi1999ram,quilis2000gone,vollmer2001ram,marcolini2003three,schulz2001multi,roediger2006ram,kronberger2008effects,jachym2009ram}. 
\citet{schulz2001multi} used SPH simulation to illustrate that a fraction of stripped gas remains in galaxy halo for about $~ 10^{8}$
yrs. The amount of gas stripped was reported to be dependent on ICM, infall velocity, halo mass, and the inclination angle. The inclination angle plays a vital role only when the central pressure of the galaxy $(P_0)$ is comparable to the ram pressure $(P_{\rm ram})$, and the disc is stripped at any inclination for $P_{\rm ram} > P_{0}$ \citep{marcolini2003three}. Ram pressure may also lead to a temporary increase in the central gas surface density for an inclined galaxy. Re-accretion of gas is also possible if the inclination angle is less than
$20^{\circ}$ \citep{vollmer2001ram}. 

Another factor that affects the RPS is the trajectory of the galaxy. The stripping due to ram pressure is the most efficient at the centre of the cluster, and the efficiency decreases as the distance from the cluster centre increase \citep{domainko2006enrichment, tecce2010ram}. Analytical models studying the effect of ram pressure also show that the Jellyfish galaxies with longer tails lie near the centre of the cluster \citep{jaffe2018gasp}. \citet{singh2019ram} showed that the galaxies falling radially inwards in a cluster lose their gas well before reaching the cluster centre, but some amount of gas is retained in case of a non-radial infall. \citet{domainko2006enrichment} showed
that about $10\%$ of the enrichment in the ICM is due to RPS, and the merger between the clusters also plays an important role in the enrichment. 

A realistic understanding of the dependency of RPS  by varying different parameters of the galaxy, like the orbit and the inclination angle is crucial to understanding star formation and, hence, the galaxy evolution. There are limited studies exploring orbital parameters and disc-wind angle on the RPS \citep{2014Bekki, 2022Wright, Akerman_2023}, even though these two are very important parameters of galaxy morphology. In this paper, we discuss the effect of ram pressure on the galaxy when it has a non-radial trajectory about the cluster centre. We also discuss the differences in the efficacy of ram pressure stripping in cases of face-on orientation and when the galaxy is inclined. The paper is structured as follows:
Section~\ref{sec:setup} contains the account of our model setup. The details of the \e~simulation (from where we extract our galaxy and cluster parameters) given in Section \ref{sec:esim} \& \ref{sec:cl-model}. The details of galaxies chosen for the analysis, the non-radial orbits, and disc-wind angle for an infalling galaxy (as extracted from these \e~simulations) are  discussed in Section~\ref{sec:gal_mod} \& \ref{sec:orbit}; In Section \ref{sec:rps} we describe the analytical model used to calculate the amount of gas stripped; Section \ref{sec:results} contains our results, their implications, and our conclusions are offered in Section~\ref{sec:discussion}.

\section{Model Setup}
\label{sec:setup}

To study the effect of orbital parameters on the ram pressure stripping, there are two key parameters: (i) the angle between the disc orientation vector and the velocity vector and (ii) the pericentric distance of the galaxies as they move towards inside the cluster. Both these parameters are affected by the dynamics and external effects like mergers as they move past the cluster medium. We are interested in galaxy evolution due to the ram pressure stripping and work with galaxies that are not significantly affected by other mechanisms like galactic mergers. Apart from these, the cluster assembly itself evolves as they accrete mass via mergers and hot-cold mode accretions.

\e~simulations are used in literature to study various physical properties of galaxy evolution; of particular interest to the present analysis is work by  \citet{2019Pallero}, who have used the simulation data for tracing the quenching history of cluster galaxies. We have used galaxy catalogue data from \e~simulation
\citep{schaye2015eagle, 2016McAlpine}, and it becomes really difficult to study the galaxy's evolution by only one process. In this work, we try to extract the trajectories that represent `smooth' in-fall. These are the trajectories that are not significantly affected by close interactions. To maximise the points in the trajectories (snapshots while falling towards clusters), we restrict
ourselves to clusters with total mass, $M_{\rm tot}\geq 10^{14} M_{\odot}$. There are six such clusters with $GroupID$ as $28000000000000-6$ with different Mass Assembly Histories (MAH). To trace the MAH of a cluster, we follow the evolution of the central galaxy ($SubGroupNum=0$) at $z=0$. The merger history of member galaxies ($SubGroupNum>0$ at $z=0$) and their distance from the central galaxy at each snapshot are calculated. We discuss the cluster selection and galaxy selections in the following subsections.

To disentangle the effects of gas depletion due to ram pressure stripping and star formation, we take the gas projections at the beginning of the infall and calculate the ram pressure stripping and do not use any proxy for gas depletion due to star formation. In this approach, we will always underestimate gas depletion as we are missing the depletion due to star formation. Therefore, the results obtained shed light on the efficiency of ram pressure stripping as a star formation quenching mechanism.

\subsection{\e ~ simulation}
\label{sec:esim}

The Evolution and Assembly of Galaxies and their Environments (\e) is a simulation suite that is a set of cosmological hydrodynamic simulations using modified Gadget3 code. It reproduces the key observational datasets, which are very resourceful for understanding the evolution of the galaxies \citep{2016McAlpine, schaye2015eagle, crain2015eagle, jiang2014n}. The simulation assumes $\Lambda$CDM cosmology with $\Omega_m = 0.307,\, \Omega_{b} = 0.04825,\, \Omega_{\Lambda} = 0.693,\, \sigma_{8} =
0.8288,\, H_0 = 67.77\, {\rm km\, s^{-1}Mpc^{-1}}\, ({\rm i.e.}\, h = 0.6777),\, n_s = 0.9611\, {\rm and}\, Y = 0.248$ \citep{2014Planck}. We use the RefL0100N1504 reference model with co-moving box length of 100 $h^{-1}$ Mpc having initial total number of particles $2 \times 1504^3$, dark matter particle mass of $9.70 \times 10^{6} M_{\odot}$ and gas-particle mass ($ m_{\rm g}$) of $1.81
\times 10^{6} M_{\odot}$.  
  
\subsubsection{Physical parameters}

Star formation in the \e~simulations is implemented using the Kennicutt-Schmidt relation \citep{1998Kennicutt}. Star formation rate decreases above the particle number density threshold depending on metallicity given by \citet{2004Schaye} as:  
\begin{equation}
n_{\rm H}^* = 10^{-1} {\rm cm}^{-3} \left(\frac{Z}{0.002}\right)^{-0.64},
\end{equation}
\noindent
where, {\it Z} is the gas metallicity. The stellar particles are generated to follow the Chabrier Initial Mass Function (IMF). The gas is prevented from artificial fragmentation by setting a density-dependent temperature floor ($T_{\rm eos} ( {\rho}_{\rm g} )$) with the equation of state given by $P_{\rm eos} \propto {\rho}_{\rm g}^{{\gamma}_{\rm eos}}$ with ${\gamma}_{\rm eos} = 4/3$, where $\rho_{\rm g}$ is the gas density, $P$ are the pressure, and $\gamma_{\rm eos}$ is the ratio of the heat capacities.  

In order to mimic gas in the cold phase, the temperature floor is normalised to $T_{\rm eos}=8000~K$ at particle number density $n_{H}=0.1 \ \mathrm{cm^{-3}}$. Gas particles with a density of more than $n_{\rm H}^*$ and the temperature $\log_{10} (T/{K}) < \log_{10} (T_{\rm eos}/{K}) + 0.5$ are assigned a star formation rate (SFR) \citep{2010Schaye, crain2015eagle}:
\begin{equation}
\label{eq:sfr}
{\dot m}_* = m_{\rm g} \ A \ (1 {\rm M}_{\odot} {\rm pc}^{-2})^{-n} \left(\frac{\gamma}{G} 
f_{\rm g} P\right)^{(n-1)/2},
\end{equation} 
\noindent
where, $\gamma = 5/3$ is the ratio of specific heats,  $f_{\rm g}$ is the gas mass fraction (kept at 1.0 for the RefL100N1504 model), and $G$ is the gravitational constant. Parameters $n=1.4$ and $A= 1.515 \times 10^{-4} \ M_{\odot} \ {\rm yr}^{-1} \ {\rm kpc}^{-2}$ are obtained from the fitting to observational data \citep{1998Kennicutt}. The radiative heating and cooling rates are computed on an element-to-element basis assuming a \citep{2001Haardt, 2009Wiersma1} ionising background and the cosmic microwave background.

The sub-grid recipe used for feedback includes supernovae  \citep[both core-collapse and type IA;][]{2009Wiersma}, active galactic nuclei \citep[AGN;][]{schaye2015eagle}, and stellar winds from Asymptotic Giant Branch (AGB) stars. The prompt stellar feedback is similar to \citet{2012Dalla}, referred to as the `stochastic feedback' model. Core-collapse supernovae are assumed to release a fraction of energy ($f_{\rm th}$) given by: 
\begin{equation}
\label{eq:fth}
f_{\rm th} = f_{\rm th,min} + 
\frac{f_{\rm th,max}-f_{\rm th,min}}{1 + \left( \frac{Z}{0.1 Z_{\odot}} \right) ^{n_Z} \left(\frac{n_{\rm H, birth}}{n_{\rm H,0}} \right) ^{-n_n}}, 
\end{equation}
\noindent
where, $n_{\rm H, birth}$ is the density of the parent gas particle, and $f_{\rm th,min}$ and $f_{\rm th,max}$ are the asymptotic values of $f_{\rm th}$. The $n_{\rm H,0}$, $n_n$, and $n_Z$ are free parameters. This released energy is injected into the neighbouring cells with a delay of 30 Myr from the birth of the stellar particle. The change in the thermal energy due to ejection from a star is isotropically distributed stochastically, which increases the temperature by $\Delta T = 10^{7.5}~K$. 

When the mass associated with a halo becomes more than
$10^{10}M_{\odot}$, a gas particle residing at the centre of the the potential well is labelled as a seed black hole and given a mass of $10^5$ ${\it{h^{-1}}}M_{\odot}$ \citep{2005SringelDiMatteo}. The mass of the central black hole is increased by the accretion of matter, which is based on a modified Bondi-Hoyle model \citep{schaye2015eagle}. The matter accreting on the central black hole returns $0.015$ times the rest-mass energy to the surroundings thermally and increases the ambient temperature by $\Delta T = 10^{8.5}~K$. Free parameters in the feedback recipe are tuned to reproduce the galaxy stellar mass function (GSMF) and galaxy sizes at
$z = 0.0$.   

\subsubsection{Galaxy catalogue and outputs} 
\label{sec:Propsec}

The outputs from the RefL100N1504 model are stored in $29$ snapshots at different redshifts between $20$ and $0$.
The dark matter halos are identified by applying the
friends-of-friends (FOF) algorithm \citep{1984Einasto}. The dark matter particles closer than $0.2$ times the mean inter-particle separation is assigned to a single halo. The subhalos are found using the SUBFIND algorithm \citep{2001Springel,2009Dolag}. Each halo in the
FOF halo is identified by a GroupID, and subhalos inside it are given a SubhaloID. We refer the reader to \citet{schaye2015eagle} and \citet{2016McAlpine} for the details of the merger tree identification and available properties for each structure. The subhalo that has the
deepest potential well is defined as the central galaxy and allocated sub-group number (SubGroupNumber) equal to zero. All the other subhalos are labelled as satellite galaxies and have SubGroupNumber equal to a positive integer value.

\subsection{Cluster Model}
\label{sec:cl-model}

As discussed at the beginning of section \ref{sec:setup}, we selected the most massive six clusters with $GroupID$ as
28000000000000-6. Figure \ref{fig:MAH} shows the mass assembly history of these clusters traced by the following central galaxy at the lowest redshift. Big fluctuations in the MAH indicate merger events. We observe that the groups 28000000000004 and 28000000000006 have had recent mergers within the last 1 Gyr; as a result, the trajectories of
galaxies are observed to be complicated. The rest of the clusters show smooth growth of mass within the last $1-2~ Gyrs$ possibility of virialization. Therefore, we restrict our analysis only to these remaining four clusters.  

We calculate the progenitor halo properties of each FOF cluster at redshift $z=0$ by tracing the properties of the parent FOF halo of central galaxy. The mass of the cluster ($M_{clus,z}$) at a redshift $z$ is taken from the \e ~ catalogues {\bf given by the entry $FOF_{200,\rm crit}$}. The radial distribution of gas inside the halo at each snapshot is assumed to follow the beta profile
\citep{king1962structure}. 
\begin{equation}
    \rho_{\beta}(R_{\rm c})= \frac{\rho_{\rm
        g0}}{\left(1+\left(\frac{R_{\rm c}}{R_{\rm
          core}}\right)^2\right)^{\frac{3\beta}{2}}}, 
\end{equation}
$R_{\rm c}$ is the cluster centric radius and $R_{\rm core}$ is the the core radius of the cluster. The core radius is considered to be, $R_{\rm core} = R_{\rm 200}/10$  \citep{2008font}, where $R_{\rm 200}$ is taken from the \e~catalog for cluster at different redshifts. The value of beta is taken to be equal to $0.6$ \citep{ota2002x,ota2004uniform,singh2019ram}.
.

\begin{figure*}
    \centering 
    \includegraphics[width=\linewidth]{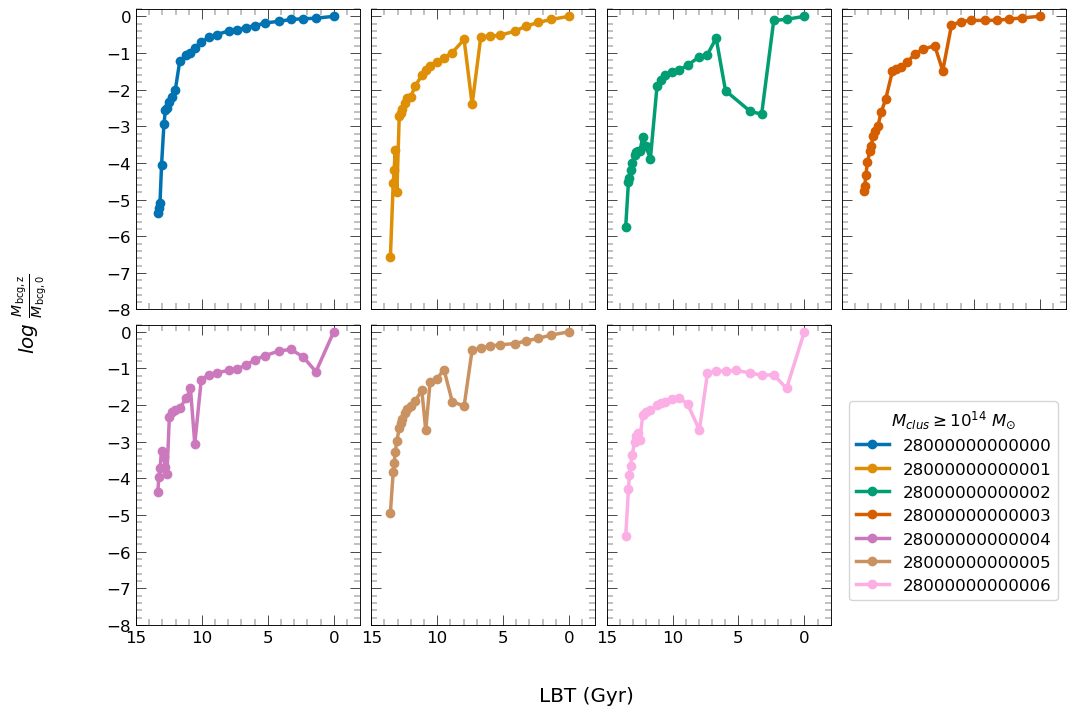}
    \caption{The mass assembly history of the six clusters. The cluster history is calculated by following the FOF halo of the central galaxy at different redshifts. Clusters 28000000000004 and 28000000000006 show signs of recent mergers within 1 Gyr and are therefore not included in the analysis.} 
    \label{fig:MAH}
\end{figure*}

\begin{figure}
    \centering 
    \includegraphics[width=0.9\columnwidth]{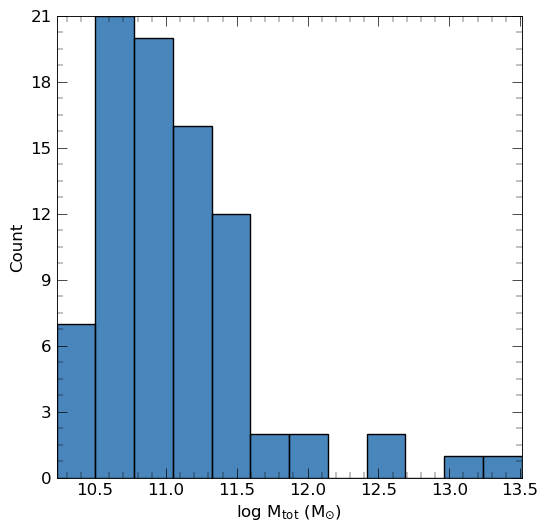}
    \caption{Distribution of total mass ($M_{\rm tot}$) of galaxies considered for the analysis. Most of the galaxies are between $10^{10}M_{\odot}$ and $10^{11.5}M_{\odot}$.} 
    \label{fig:mdist}
\end{figure}

\begin{figure}
    \centering
    \includegraphics[scale=0.5]{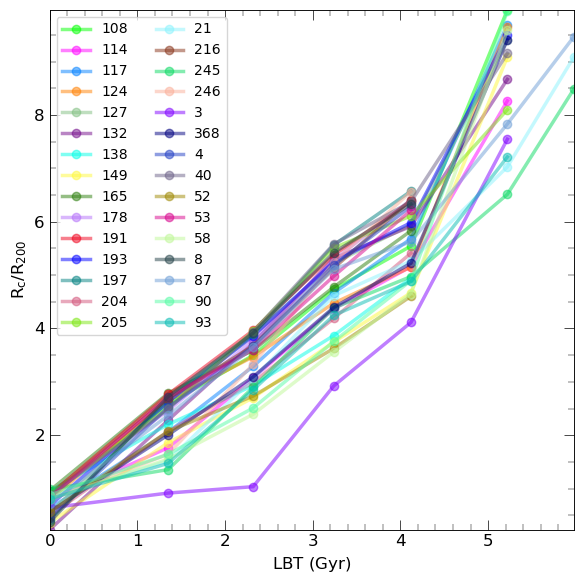}
    \caption{Distance to the centre ($\mathrm{R_c/R_{200}}$) plotted with lookback time (LBT; in Gyr) for different subhalos in the cluster $28000000000002$. The centre is marked by the black point around which the orbit turns. The three-position axes have been normalised with the virial radius $(R_{\rm 200})$.} 
    \label{fig:trajec}
\end{figure}

\begin{figure}
    \centering
    \includegraphics[width=0.9\columnwidth]{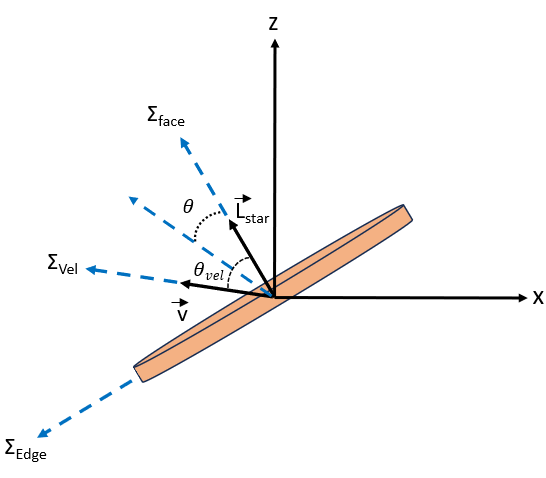}
    \caption{Inclinations of disc considered in the analysis. Vectors $\vec{v}$ and $\vec{L_{star}}$ denote the velocity and stellar angular momentum of the galaxy, respectively. $\theta$ denotions the inclination angle of velocity with respect to angular momentum axis. We consider three projections of Face ($\theta=0^{\circ}$), Edge ($\theta=90^{\circ}$), and Vel ($\theta=\theta_{\rm vel}$) corresponding to the projection along angular momentum direction, edge of the disc (perpendicular to angular momentum direction) and along velocity direction respectively.} 
    \label{fig:incl}
\end{figure}

The central density value ($\rho_{\rm g0}$) in the beta profile used in our model is $\mathrm{3.53e^{-27} \ \rm g/cm^3}$, and it has been calculated using the values from Navarro, Frenk, and White (NFW) Profile that the dark matter follows \citep{navarro1997universal}.  
\begin{equation}
    \frac{\rho(R_{\rm c})}{\rho_{\rm
        crit}}=\frac{\delta_c}{\left(\frac{R_{\rm
          c}}{R_s}\right)\left(1+\frac{R_{\rm c}}{R_s}\right)^2}, 
\end{equation}
$\rho_{\rm crit}$ is the critical density and is equal to $\rm 3H^2/(8\pi G)$, $R_s (=R_{\rm 200}/c_{\rm 200}$) is the scale radius, and $\delta_c$ is the characteristic density and is a dimensionless quantity given by: 
\begin{equation}
    \delta_c=\frac{200}{3}\frac{c^3}{ln(1+c)-\left(\frac{c}{1+c}\right)},
\end{equation}
\noindent
where $c$ is the concentration of the halo, calculated as:
\begin{equation}
	c = \frac{6.71}{\,\,\,\,(1+z)^{0.44}}
      \left(\frac{M_{clus,z}}{2\times10^{12}h^{-1}M_{\odot}}\right)^{-0.092},  
\end{equation}
where $M_{clus,z}$ is the total mass of the cluster at redshift $z$. For cosmological parameters, we use the results from \cite{2014Planck}.

\subsection{Galaxy selection}
\label{sec:gal_mod}

We extracted all the parameters of galaxies in our semi-analytic simulation from the \e-database. We only select the member galaxies ($SubGroupNum>0$) that reach a cluster-centric distance ($R_{\rm c}$) less than one virial radius ($R_{\rm 200}$). The parameters include the GalaxyIDs, velocity, and position at different redshifts, the distance of the galaxy from the cluster centre, and their masses. We
select the galaxies with initial gas mass (at the start of infall) $M_{\rm ini, gas}\geq 10^{8} M_{\odot}$.  We use the \e particle data \citep{schaye2015eagle, 2016McAlpine, crain2015eagle} to make projections in a 2D grid of stellar particles and gas particles at the start of infall. We calculate the initial gaseous and stellar surface densities with a resolution of 1 kpc. While projecting the gas, we take only the cold gas ($T<10^4~K$) in the galaxy.  

\subsection{Galaxy Orbit}
\label{sec:orbit}

Merger history for each member galaxy in clusters is calculated. The distance of the centre of the progenitors of member galaxies from the centre of the cluster progenitor ($R_{\rm c}$) is calculated at each redshift. We focus on the galaxies falling inward for the first time. As galaxies fall towards the centre of a cluster, they might
experience encounters with other galaxies. The encounter can result in a change in galaxy trajectory, and a change in local effects of interaction could dictate galaxy evolution. To focus only on the effects of trajectory and evolution by ram pressure stripping, we select galaxies for which the distance to the centre of the cluster ($R_{c}$) is monotonically decreasing with redshift. We identify 84
such unique trajectories in the five clusters selected for analysis. While trajectories can be expected to take galaxies away from the centre, this is not expected if we focus on the first infall.

In Figure \ref{fig:mdist} we show the distribution of the total mass of all particles ($M_{\rm tot}$) taken from the subhalo catalogue of galaxies considered for analysis. Most of the galaxies are in the range of $10^{10}M_{\odot}$ to $10^{11.5}M_{\odot}$. Figure \ref{fig:trajec} shows how the distance to the centre decreases with lookback time (LBT)  for the galaxies considered in the cluster 28000000000002. The points on the line indicate the snapshots of galaxies. All galaxy trajectories converge monotonically towards the centre of the cluster. We get all such trajectories for all the clusters considered for analysis (described in the previous section).

\subsubsection{Inclination Angle}
\label{sec:incli}

We consider three projections of the galaxy for calculating the ram pressure stripping (shown in Figure~\ref{fig:incl}): 

\begin{itemize}
\item {\bf Face}: The galaxy gas and stellar mass density projected along the angular momentum axis of the stellar disk  ($\theta= 0^{\circ}$). 
\item {\bf Edge}: The galaxy gas and stellar mass density projected along the edge of the stellar disk  ($\theta= 90^{\circ}$). 
\item {\bf Vel}: The galaxy gas and stellar mass density along the velocity vector of the galaxy ($\theta= \theta_{\rm vel}$). 
\end{itemize}

We calculate these projections at the start of infall and apply the ram pressure stripping at each redshift to every bin of 2D grid. The value of the three-dimensional centre of mass position and the velocity vector for the infalling subhalo, as well as the central subhalo is calculated from the \e dataset.  The angle of inclination ($\theta$) which is the angle between the disc orientation vector and the velocity vector varies as the galaxy travels in the ICM and is calculated at each point along the trajectory.

\begin{figure*}
    \centering
    \includegraphics[scale=0.6]{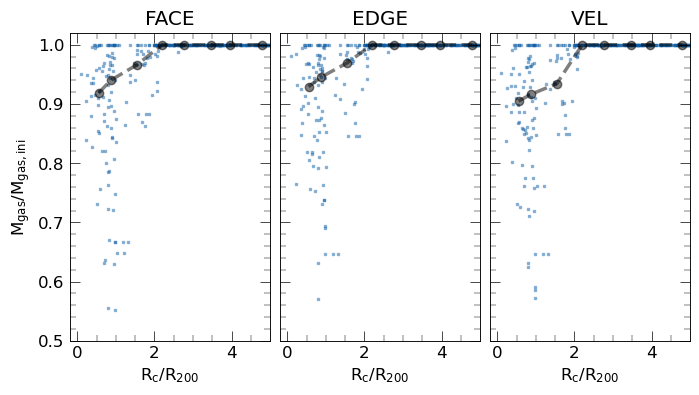}
    \caption{Fraction mass remaining ($M_{\rm gas}/M_{\rm gas,ini}$) as a function of $R_{\rm c}/R_{\rm 200}$ for all the galaxies considered by summing up all the surviving gas pixels in 2D grid at each snapshot. The three subplots show the results for the face, edge and velocity projections. The colour line indicates the median fraction of gas removed for a given $R_{\rm c}/R_{\rm 200}$. The ram pressure stripping is not effective at a distance away from central galaxies;  therefore, the fraction is close to one. For galaxies close to the centre, the fraction of mass removed increases depending on the projection angle.} 
    \label{fig:rps}
\end{figure*}

\subsection{Ram Pressure Stripping}
\label{sec:rps}

The method adopted in this paper to evaluate the RPS is similar to \citet{singh2019ram}. In this section, we give a brief account of how stripping is calculated. The ram pressure force pushes the gas in a direction opposite to the velocity of the infalling galaxy, but the gas is stripped only when due to ram pressure, the force per unit area $F_{\rm ram}\,=\,\rho(R_{\rm c})v^2(R_{\rm c})$, is strong enough to exceed the gravitational restoring force, $F_{\rm res}\,=\,2\pi G \Sigma_{\rm g}(R) \Sigma_{\rm s}(R)$. The restoring force retains the gas and prevents it from escaping the galactic disc. Here, $\rho(R_{\rm c})$ is the density of the ambient medium, $v(R_{\rm c})$ is the velocity with which the galaxy moves through the medium,
$R_{\rm c}$ is the cluster centric radius, $G$ is the gravitational constant, $\Sigma_{\rm g}(R)$ is the surface density of the gaseous component, $\Sigma_{\rm s}(R)$ is the surface density of the stellar component, and $R$ is the disc radius of the galaxy. Thus, the condition for the RPS to be effective is \citep{gunn1972infall}: 

\begin{equation}
    \rho(R_{\rm c})v^2(R_{\rm c})>2\pi G\Sigma_{\rm g}(R)\Sigma_{\rm s}(R).
\end{equation}

To understand the effect of the inclination angle on the stripping, we consider the acceleration due to ram pressure. The force of ram pressure is a vector quantity; therefore, the component perpendicular to the galactic plane acts as the effective ram pressure on the galaxy for stripping of gas. The parallel component is responsible for compressing the gas \citep{singh2019ram,troncoso2020better}. This introduces a factor of $\sin \theta$ with $F_{\rm ram}$, where $\theta$ is the angle of inclination as discussed in section~\ref{sec:incli}. The acceleration on the galactic disc is in the opposite direction of its infall
velocity and is calculated as: 
\begin{equation}
    a(R_{\rm c},R)=\frac{F_{\rm ram}(R_{\rm c})\sin\theta-F_{\rm res}(R)}{\Sigma_{\rm g}(R)}. 
\label{eq:acceleration}
\end{equation}
From the above equation, it is clear that the ram pressure is reduced if the galaxy is inclined \citep{marcolini2003three, sharma2012structure, vollmer2001ram}. Therefore, the ram pressure is expected to be in decreasing order in the case of Face, Vel, and Edge projections are discussed in the previous section.  

For our analysis, we consider the ram pressure stripping in each bin for three projections of the galaxy. We consider the gas in the bin to be stripped if the acceleration defined in equation~\ref{eq:acceleration} is positive. This is different from the method used in \citet{singh2019ram} as the trajectory was calculated analytically. The gas annulus was considered stripped only if it had reached a constant distance above the galactic disc. Since, in the present analysis, the trajectory is calculated using discrete snapshots from the simulation, such a condition is not possible.

\section{RPS Results}
\label{sec:results}

In the following subsections, we explore the effect of different parameters on the ram pressure stripping of galaxies moving inside the cluster. In Section \ref{sec:res_incl} we study the effect of orientation of galaxies followed by galaxy total mass in Section
\ref{sec:res_mass}. We discuss how the impact parameter affects the fraction of mass removed in Section \ref{sec:res_impact} followed by the discussion in Section \ref{sec:discussion}. 

\subsection{Effect of inclination angle}
\label{sec:res_incl}

We begin the numerical exploration of ram pressure by taking the value of $R_{\rm c}/R_{\rm 200}$ taken from the orbits given in section~\ref{sec:orbit}. In Figure \ref{fig:rps}, we show the fraction of mass remaining ($M_{\rm gas}/M_{\rm gas,ini}$) as a function of $R_{\rm c}/R_{\rm 200}$. The remaining mass of gas is calculated at every snapshot in the orbit by summing up all the surviving gas pixels in a 2D grid. The line indicates the median fraction of gas removed for a given $R_{\rm c}/R_{\rm 200}$. At the distances of $R_{\rm c}/R_{\rm 200}>2$, the fraction of mass remaining is close to one, indicating the ram pressure is not effective in
removing the gas. Within two virial radii of the cluster, the ram pressure decreases the gas mass depending on the projection angle.  

In Figure \ref{fig:mrem_kde} we show the distribution of the fraction of gas mass removed for the three orientations. The peak of all three distributions is between $0-3$\%. The vertical lines show the location
of the median for the three distributions. The median gas removed is more in the face-on orientation and minimum in the edge-on orientation. This result is in agreement with our understanding of the effect of orientation described in section~\ref{sec:rps}. We note that although different, the median values are very close to each other, and
there is no significant difference between the three
projections. Inclination does not have much effect on the amount of gas stripped.

\begin{figure}
    \centering
    \includegraphics[width=0.9\columnwidth]{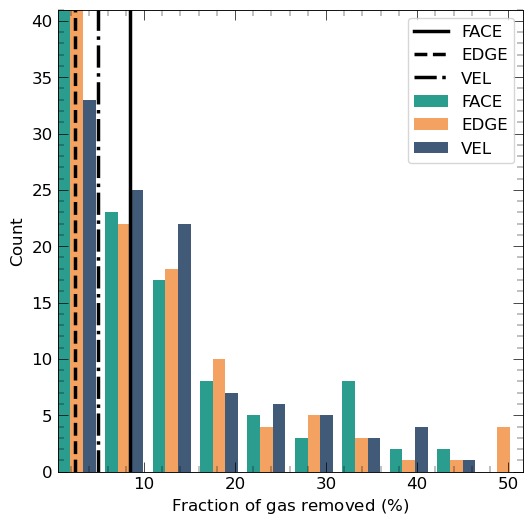}
    \caption{Histogram of the fraction of gas removed for all the galaxies considered. The three colours represent the three projections considered for the analysis. Vertical lines show the median values of the fraction of gas removed. The median fraction of mass removed is more in the face-on orientation and minimum in the edge-on orientation.} 
    \label{fig:mrem_kde}
\end{figure}

\subsection{Effect of mass of galaxy}
\label{sec:res_mass}

In \cite{singh2019ram} we observed that the total mass of a galaxy is important in preserving the gas mass while moving inside a cluster environment. Massive galaxies are expected to have higher stellar mass surface density per bin and should be expected to have a higher restoring force. In Figure \ref{fig:rps_mass} we show the trajectory
of galaxies in $M_{\rm gas}/M_{\rm gas,ini}-R_{\rm c}/R_{\rm 200}$ plane for the highest and lowest mass galaxies in our sample for all projections. 

The galaxy with the highest total mass retains all its gas up to one virial radius. The retention in the case of face-on projection is higher compared to the edge-on case. This can be attributed to the galaxy having a non-discy morphology. However, in the case of the lowest mass galaxy, the stripping is more effective in the face-on projection compared to the edge-on projection. The stripping starts to be effective beyond two virial radii.

In Figure \ref{fig:proj_mass} we bin the total mass of galaxies ($M_{\rm tot}$) and explore the fraction of gas remaining ($M_{\rm gas}/M_{\rm gas,ini}$) for different projections. The last bin is wider due to the smaller number of high-mass galaxies in our sample. Note that the colours and the number indicate the median fraction of mass removed. For the galaxies with total mass less $M_{\rm tot}<10^{10.5} \ M_{\odot}$, the effect of RPS is more pronounced, resulting in more than 20\% more gas loss. As observed in the previous section, the face-on case is more effective in removing gas than the edge-on case.

\subsection{Effect of impact parameter}
\label{sec:res_impact}

The impact parameter (denoted by $b$) of a galaxy is the perpendicular distance to the trajectory of a galaxy moving inside a central field. A low impact parameter would indicate that the galaxy will pass close to the centre, thereby experiencing a higher ram pressure. The
impact parameter of the galaxy is approximated by taking the velocity at the last snapshot ($z=0$). In Figure \ref{fig:hist_impact} we show the distribution of normalised impact parameters ($b_{\rm min}/R_{\rm 200}\sim 1$) for all the galaxies considered. The impact
parameters show a uniform distribution between minimum and maximum values. 

In Figure \ref{fig:rps_impact} we show the RPS during the  trajectory of galaxies in $M_{\rm gas}/M_{\rm gas,ini}-R_{\rm c}/R_{\rm 200}$ plane for different projections. We note that this might not be a good approximation for the galaxies beyond the virial radius. For high impact parameters ($b_{\rm min}/R_{\rm  200}\sim 1$) the ram pressure is not enough for galaxies to lose any gas. We observe a similar trajectory for all the projections with curves overlapping each other. In the case of low-impact parameter ($b_{\rm min}/R_{\rm 200}\ll 1$) the galaxies experience a close encounter and may lose up to 25\% of their gas mass. The stripping is more in the case of Face-on projection. 

To explore the trend for other values of impact parameters in Figure \ref{fig:proj_impact}, we show the fraction of mass remaining ($M_{\rm gas}/M_{\rm gas,ini}$) for different bins of impact parameters ($b_{\rm min}/R_{\rm 200}$) and projections. Galaxies on trajectories with low-impact parameters ($b/R_{\rm 200}\sim 0-0.2$) can have $10$\% more gas removed compared to high-impact parameters trajectories. The effect is similar for all the projections.

\subsection{Contribution of RPS}
\label{sec:comp}

In Figure \ref{fig:sim_com}(left) we show the comparison between the gas remaining in the simulation ($M_{\rm sim}$) and compare it with our analytical results ($M_{\rm ana}$) for RPS of gas in disc at the last snapshot. The galaxies that completely lose gas are given mass $10^{7}\ M_{\odot}$. The dashed line shows the equal masses. We observe that the remaining gas matches our result only for a few galaxies, and for most of the galaxies, the gas loss can not be explained only by ram pressure stripping.    

To compare the contribution of ram pressure stripping to the gas depletion of galaxies, we calculate the contribution ratio as: 
\begin{equation}
    \mathrm{RPS \ contribution}(\%)  = \frac{\rm Gas \ depleted \ in
      \ analytical \ model}{\rm Gas \ depleted \ in \ simulation}\times
    100 
    \label{eq:comp}
\end{equation}
\noindent
Figure \ref{fig:sim_com}(right) shows that for the majority of the galaxies, the RPS contribution is less than $10$\%. The value does not change with the projection angle. This indicates that the ram pressure stripping is not a major process for the gas depletion of the galaxies
within the cluster environment. 

We note that the method presented in this paper only discusses the effect of RPS on the gas in the disc. The method does not take into account the hot gas residing in the circumgalactic medium (CGM). Processes like AGN and supernova feedback can lead to gas thrown out of the CGM, resulting in quenching of star formation due to no fresh supply of cold gas from surrounding, referred to as `starvation' \cite[e.g. ][]{2012Fabian,2015King,2020Trussler}. Additionally, hot gas surrounding a galaxy (like in ICM) can cut-off the fresh supply of gas to the galaxy, resulting in quenching of star formation, referred to as `strangulation' \cite[e.g. ][]{1980Larson,2008van,2020Trussler}. In a hydrodynamical simulation, these processes can play a crucial role in regulating the rate of star formation.

\begin{figure}
    \centering
    \includegraphics[width=0.9\columnwidth]{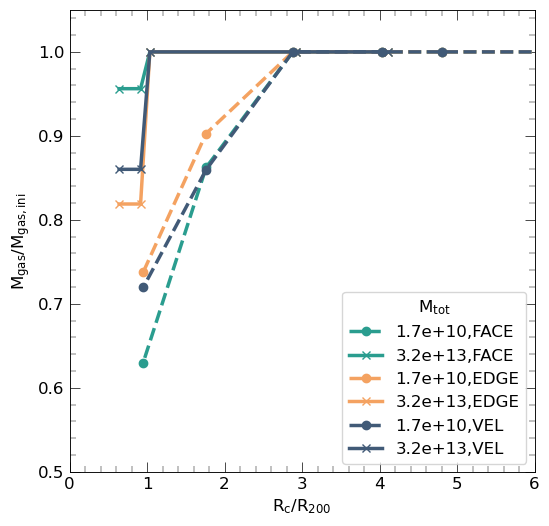}
    \caption{The fraction of mass remaining ($M_{\rm gas}/M_{\rm gas,ini}$) as a function of $R_{\rm c}/R_{\rm 200}$ for all the galaxies considered. Different lines show the trajectories of the most massive and least massive galaxies for the three projections. The least massive galaxy is not able to retain its gas as it moves through the cluster within $R_{\rm c}/R_{\rm 200}<3$. The most massive galaxy, on the other hand, can retain their gas and does not lose more than 20\% of their mass for the most effective face-on projection.} 
    \label{fig:rps_mass}
\end{figure}

\begin{figure}
  \centering
  \includegraphics[width=0.9\columnwidth]{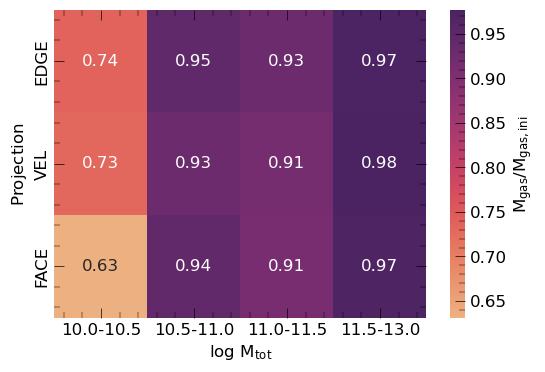}
  \caption{The fraction of mass remaining ($M_{\rm gas}/M_{\rm gas,ini}$) for different total mass of galaxies ($M_{\rm tot}$) and different projections. The colours and the number indicate the median fraction of mass remaining. For the galaxies with total mass less $M_{\rm tot}<10^{10.5} \ M_{\odot}$ the effect is more in the face-on projection, which can lose more than 30\% of their mass.} 
  \label{fig:proj_mass}
\end{figure}

\begin{figure}
    \centering
    \includegraphics[width=0.9\columnwidth]{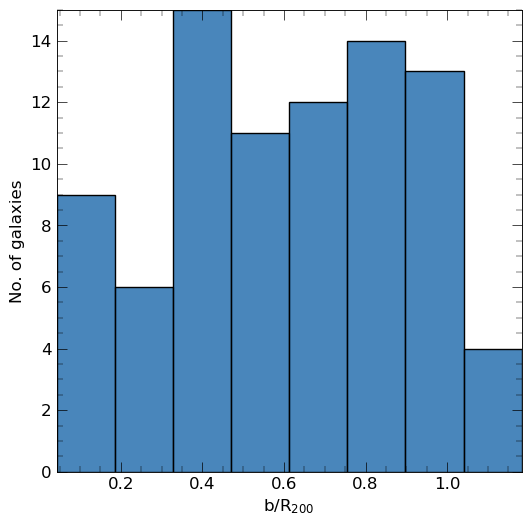}
    \caption{Distribution of impact parameter ($b$) normalized by cluster radius ($R_{\rm 200}$) for all the galaxies considered in the analysis. The impact parameters show a uniform distribution between the minimum and maximum values of the impact parameter.} 
    \label{fig:hist_impact}
\end{figure}

\begin{figure}
    \centering
    \includegraphics[width=0.9\columnwidth]{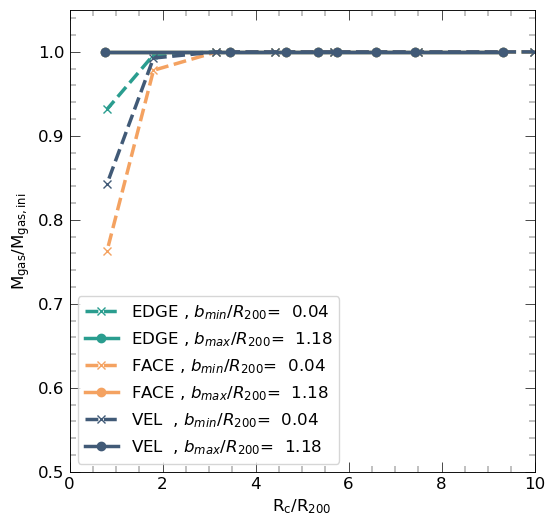}
    \caption{Same as Figure \ref{fig:rps_mass} but for different impact parameters. All curves for the case of highest impact parameter overlap in the plot. Two cases show the highest and lowest impact parameters cases for the three projections. The gas is retained for all the projections in the case of a high impact parameter. For the low-impact parameter case, more gas is retained in the edge-on case than in the face-on projection.} 
    \label{fig:rps_impact}
\end{figure}

\begin{figure}
  \centering
  \includegraphics[width=0.9\columnwidth]{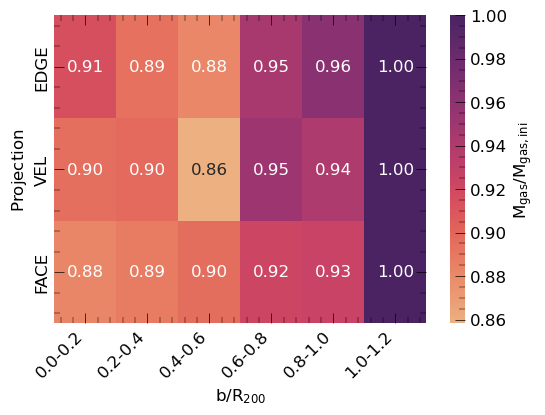}
  \caption{The fraction of mass remaining ($M_{\rm gas}/M_{\rm gas,ini}$) for different impact parameters ($b/R_{\rm 200}$) and different projections. The colours and the number indicate the median fraction of mass remaining. Galaxies with low-impact parameters ($0-0.2$) can have $10$\% more gas removed compared to high-impact parameters. The effect is similar for all the projections.} 
  \label{fig:proj_impact}
\end{figure}

\begin{figure*}
    \centering
    \includegraphics[scale=0.5]{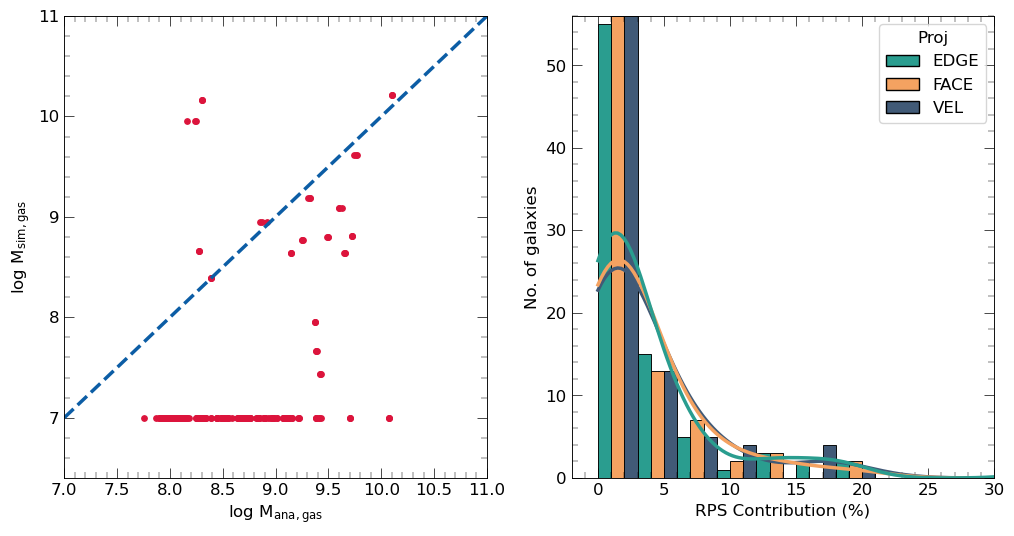}
    \caption{(left)Comparison between the gas lost in the simulation ($M_{\rm sim}$) and compare it with our analytical results ($M_{\rm ana}$) of RPS stripping of gas from the disc at the last snapshot. The dashed line indicates equal mass loss. The cases with no gas are indicated with the gas mass of $10^{7}\, M_{\odot}$. (right) Contribution to the gas depletion due to ram pressure stripping (see the text for details). The analytical model indicates that ram pressure stripping is not a major contributor to the depletion of the gas.} 
    \label{fig:sim_com}
\end{figure*}

\section{Discussion and conclusions}
\label{sec:discussion}

The relevance of current work in understanding the complex process of variation of star formation rate and hence the galaxy evolution is the the key question that we address now. The amount of gas present in galaxies determines the star formation, and ram pressure stripping is considered one of the main deriving forces behind removing the gas from the galaxy as it passes through the ambient medium. Studying its variation with the morphology of galaxies and the ambient medium is vital for the study of galaxy evolution. Ram pressure has been extensively studied since the work by \citet{gunn1972infall}. Previous works have already been discussed in the introduction. One of the key challenges to the RPS scenario is the discovery of galaxies near cluster centres that contain a significant amount of cold gas and are also forming stars, whereas simple estimates suggest that RPS should lead to the removal of all cold gas from all galaxies in the cluster environment.  

Work pertinent to the present paper is the work by
\citet{singh2019ram} in which they have given a semi-analytical model for RPS and discussed in detail the variation of RPS  with the ambient medium properties and galaxy mass. In that paper, we showed that inhomogeneities in the cold interstellar medium of galaxies slow down
RPS and galaxies are able to retain some gas as they traverse the ICM. Recently, \citet{2022Kuchner} have shown that galaxies falling along filaments that are feeding matter into clusters may not lose gas to RPS due to a cooler ambient environment and lower relative
velocities.   

In the present work, we focus on how the effect of ram pressure is modified if the galaxy's orientation w.r.t. the orbital plane is taken into account, along with the deviation of the orbit of an infalling galaxy from the radial infall. We also disentangle the gas depletion
due to the star formation and other processes operating inside the cluster environment. The results, therefore, act as an upper bound of the efficiency of ram pressure as the mechanism to remove gas from galaxies and thereby quenching star formation. We have used \e-simulation database to extract these properties of galaxies. Galaxies selected from the database are such that they span a range of three orders of magnitude in mass, different types of
orbits, and a vast range of variation in inclination angle as discussed in Section~\ref{sec:setup}. 

We have attempted to address the following questions in the current work: 
\begin{enumerate}
 \item  What is the effect of the inclination angle of the infalling galaxy in comparison to face-on systems on the RPS\,? 
 \item How the gas removed in the galaxies evolved using only RPS compared with their respective gas content in \e~simulations\,? 
 \item What is the effect of the environment on the efficiency of RPS as a mechanism to remove gas from galaxies\,? 
\end{enumerate}

Our model explained in section~\ref{sec:rps}, although simple, captures the essence of the above-mentioned questions quite efficiently. The ram pressure stripping depends on the trajectory, galaxy mass, and inclination angle. For the face-on galaxies, the effect of ram pressure increases as the galaxy approaches the centre. When the galaxy is inclined, the stripping process is slowed, and less gas is stripped. However, for the highest mass galaxies, the face-on galaxies seem to retain more gas, possibly due to deviation from the disc-like morphology of the galaxy. 

To understand the observed pattern better, we further compared the gas mass retained in the galaxies in \e~simulation and that obtained from our analytical model. The comparison shows that only ram pressure is not responsible for removing the gas from the galaxies in \e~simulations, and other mechanisms may dominate. A few percent decrease in the amount of gas in our analytical model could occur had we taken into account the processes like star formation etc. However, the contrast is large enough to conclude that RPS is not efficient in gas depletion in galaxies.  

The presence of other galaxies in the vicinity would contribute to the gas removal process, indicating that processes like tidal interactions, harassment, and viscous drag do contribute equally in the removal of gas from the interstellar medium of galaxies. Thus apart from inhomogeneities in the interstellar medium and the effect
discussed by \citet{2022Kuchner}, we find that the inclination of infalling galaxies contributes marginally to a reduction in the efficacy of RPS and hence explain the survival of cold gas in galaxies near centres of clusters of galaxies. Recent results by \citet{Hough-2022} also indicates that RPS alone can not quench the gas in galaxies during the first pass through the cluster centre. These new findings suggest that the traditional RPS models needs to be revisited and reanalyzed for gas quenching in the disc galaxies. 

In our analytical model, we only consider the gas loss in disk (cold gas; $T \leq 10^{4} \ K$) due to the effect of RPS and ignore the other processes like strangulation, harassment, and starvation. These processes can be crucial in quenching the star formation by cutting the supply of gas present in the CGM in the form of hot gas ($T > 10^{4} \ K$). These could lead to underestimation of quenching in our model. Therefore, the results presented in this paper should be taken as a lower bound on the efficiency of RPS and how it changes due to different orbital parameters.

This work uses the \e~simulation to perform the analysis. We focused on only the largest clusters in the simulation with total mass $M_{\rm tot}\geq10^{14} \ M_{\odot}$ in addition to limiting our analysis to the radial trajectories, leading to only 85 galaxies in the sample. The size of the simulation box limits the number of massive clusters \citep{2005Bagla,2009Bagla}. Therefore, a simulation with a larger box would have more massive clusters that can have more extreme cases of RPS due to a higher density of ICM. Using a larger box will also increase the statistical sample to give a better bound on the efficiency of RPS for different trajectories.

Our results in Section \ref{sec:comp} show that some galaxies in hydrodynamical simulation show a gas depletion that is comparable to the estimates from just RPS, while others show higher depletion. Using a higher box size simulation will allow us to segregate these galaxies with a good statistical sample size, allowing us to compare and contrast their physical and orbital properties in greater detail. We leave this to future work.

\section*{Acknowledgements}

The authors would like to thank the anonymous referee for his in-depth suggestions, which has improved the results. We acknowledge the Virgo Consortium
for making their simulation data available. The \e~simulations were
performed using the DiRAC-2 facility at Durham, managed by the ICC, and the
PRACE facility Curie based in France at TGCC, CEA, Bruy\`eres-le-Ch\^atel. The authors are thankful for the library and computational facilities at the Korea Institute for Advanced Study (KIAS Center for Advanced Computation Linux Cluster System), Thapar Institute of Engineering and Technology, Patiala, and the Indian Institute for Science Education and Research (IISER) Mohali. Ankit Singh is supported by a KIAS Individual Grant PG080901 at the Korea Institute for Advanced Study (KIAS). Mamta Gulati acknowledges Thapar Institute of Engineering and Technology, Patiala for SEED grant SEED/Money/TU/DORSP/57/3978 and DST-FIST (Govt. of India) for the grant SR/FST/MS-1/2017/13 to support this research work. This research has made use of NASA's Astrophysics Data System. 

\section*{Data Availability}

The data underlying this article will be shared on reasonable request to the corresponding author.

\bibliographystyle{mnras}
\bibliography{references} 
\label{lastpage}
\end{document}